\documentstyle[twocolumn,aps]{revtex}

\begin{document}

\title{
Giant transverse magnetoresistance in an asymmetric system of three
GaAs/AlGaAs quantum wells in a strong magnetic field at room temperature
}

\author{V.I.\,Tsebro\thanks{e-mail: tsebro@sci.lebedev.ru},
O.E.\,Omel'yanovskii, V.V.\,Kapaev, Yu.V.\,Kopaev}
\address{P.N.\,Lebedev Physics Institute, Russian Academy of Science,
117924 Moscow, Russia}
\author{V.I.\,Kadushkin}
\address{Scientific-Research Technological Institute,
390011 Ryazan', Russia}

\date{December 25, 1996}

\maketitle

\begin{abstract}
The giant transverse magnetoresistance is observed in the case of
photoinduced nonequilibrium carriers in an asymmetric undoped system of
three GaAs/AlGaAs quantum wells at room temperature.  In a magnetic field
of 75 kOe, the resistance of nanostructure being studied increases by a
factor of 1.85. The magnetoresistance depends quadratically on the
magnetic field in low fields and tends to saturation in high fields. This
phenomenon is attributed to the rearrangement of the electron wave
function in magnetic field.  Using the fact that the incoherent part of
the scattering probability for electron scattering on impurities and bulk
defects is proportional to the integral of the forth power of the envelope
wave function, the calculated field dependence of the magnetoresistance is
shown to be similar to that observed experimentally.
\end{abstract}

\pacs{}

In our previous paper~\cite{omel96} we studied for the first time
in detail the lateral photogalvanic effect (PGE) in an asymmetric
system of three GaAs/AlGaAs wells illuminated with white light of
various intensities in a strong magnetic field.  The spontaneous PGE
current $J^{PGE}$ was shown to exhibit a maximum as a function of
magnetic field $H$, the phenomenon earlier predicted theoretically in
\cite{gorba93}.  We have also found that, at room temperature, the PGE
voltage reaches the value of several tenth of Volt per millimeter of
the specimen length in the illuminated region, and exhibits only a weak
dependence on the light intensity. The temperature dependence of PGE
was found to be rather weak: $J^{PGE}$ decreases by a factor of two
upon cooling from room temperature to $\sim$~200~K.

The conclusion \cite{gorba93} about the maximum of $J^{PGE}$ vs $H$
follows from an analysis of the expression for the toroidal moment density
$\bf T$, to which the spontaneous PGE current is proportional.
On the other hand, we have also proposed \cite{omel96} that such a maximum
may be caused by a strong transverse magnetoresistance (TMR) of the
nanostructure. Therefore, it was of interest to measure the TMR and its
field dependence, particularly at high, room temperature, where the PGE
was shown to be the largest.

In the present paper, we report the results of investigation of the TMR in
the asymmetric system of three quantum wells, whose structure
(see Fig.~1) was exactly the same as in our previous
paper~\cite{omel96}. Namely, the samples of the
$i$-Al$_x$Ga$_{1-x}$As/$i$-GaAs ($x$=0.25) nanostructures
containing three quantum wells with layer of width $L_W$ = 54, 60 and 70~\AA\
separated by barrier layers of width $L_B$ = 20 and 30 \AA \ were
investigated. This asymmetric system of tunneling-coupled quantum wells
was sandwiched between two wide (200~\AA) $i$-Al$_x$Ga$_{1-x}$As
($x$=0.25) barriers layers adjacent to an $i$-GaAs (1~$\mu$m) buffer layer
and to an $i$-GaAs (200~\AA) layer covering the structure.

The samples were rectangular with dimensions of the order of 8$\times
2$~mm with single pair of in-line contacts (1, 2 in Fig.~1). The
contacts were produced by the allowing in of indium. The measurements
were carried out at room temperature in a specially designed
``warm-field'' insert of a superconducting solenoid. Light from a
halogen lamp was delivered to the sample along a flexible optical
fiber. The maximum power of the radiation delivered to the sample was
of the order of 5~mW. The contacts and the adjacent parts of the
samples were covered with a special shield (3), so that only the
central part of the sample was illuminated. The samples were oriented
with the plane of the layers parallel to the magnetic field and with
the line of the contacts perpendicular to the magnetic field.

The measurement circuit shown in Fig.~1 was a simple closed one with the
sample connected in series with a source of controllable DC bias voltage
$E_v$ and a standard measuring resistance $R_n$. The current $J$ in the
circuit was determined from the voltage drop across the $R_n$. Note that
the measured current was essentially the short-circuit current $J_{sc}$,
since the resistance of the samples (of order of 100~M$\Omega$ under
nominal illumination) was much larger than $R_n$ (10~k$\Omega$).
During the experiments, the magnetic field dependencies $J(H)$ were
measured at different $E_v$ and nominal fixed illumination.
As the magnetic field was scanned bidirectionally from $-$75~kOe to
75~kOe, the measured values of $J$ were stored and averaged over a large
number of readings.

Figure~2 shows the magnetic field dependencies $J(H)$ measured at
different bias voltages $-6$~V $< E_v < 6$~V. The odd curve
$J(H)$ (at $E_v = 0$) represents the magnetic field dependence of the
spontaneous PGE current $J^{PGE}(H)$, that was investigated in details
earlier in paper~\cite{omel96}. As positive or negative bias voltage is
applied, the $J(H)$ curves are shifted up or down. At the same time, a
strong decrease of the absolute values of the current with increasing
magnetic field is observed, i.e. a high magnetoresistance
becomes apparent.

The $J(E_v)$ values at given $H$ enable us to judge to what extent the
current--voltage characteristics of the samples are linear and symmetric
in so wide range of bias voltage. The data presented in Fig.~2 reveal
that the current--voltage characteristics are linear and
symmetric with the accuracy to within several percents throughout
the studied $E_v$ and $H$ ranges. Special tests showed that the degree
of nonlinearity and asymmetry primarily depends on the quality
(symmetry) of the contacts and show a noticeably tend to reduction
with increasing magnetic field.

Since, contrary to PGE, the magnetoresistance is an even function of $H$,
the procedure of obtaining its magnetic field dependence and excluding the
contribution of PGE consists in subtracting the $J(H)$ dependence
measured at negative $E_v$ from that measured at positive $E_v$:
$R(H) = 2E_v/(J^+(H)-J^-(H))$.  The magnetic field dependencies of the
magnetoresistance obtained using this procedure for both low (1~V) and
high (6~V) absolute values of $E_v$ are shown in Fig.~3 as $\Delta
R(H)/R(0)$. A small difference between these two curves can be related to
a weak field-dependent nonlinearity of the current--voltage
characteristics. In low magnetic fields ($H <\ $10~kOe), the
magnetoresistance is a quadratic function of $H$. In high fields, a
tendency to saturation is seen. In a maximum field of 75~kOe used in
present measurements, the resistance of the nanostructure increased by a
factor 1.85.

For interpretation of the data obtained, calculation of the energy
spectrum and the wave functions for the studied asymmetric nanostructure
has been performed by the envelope method in a magnetic field normal to
the plane of the sample ($x$-axis). Figure~4 (curves 1--3) shows the
magnetic field dependencies of the probabilities for an electron to be
located in the corresponding (according to the number in Fig.~1) quantum
well in the minimum of the first conduction subband of the spatial
quantization.  Note that the nanostructure studied was specially designed
in such a way that, in zero magnetic field, the probabilities for
electrons to be located in the ground state in the narrower quantum wells
(1 and 2) are rather high. As seen from Fig.~4 (curve 3), the
magnetic field induced rearrangement of the wave function leads to the
localization of electrons in the widest well that certainly affects
the conductivity (resistivity) of the nanostructure in the lateral
direction.

The resistivity of the nanostructure in the lateral direction is
determined by the electron scattering on the heterojunction imperfections,
residual impurities, and bulk defects.
The localization of the wave function in the center of the widest well is
likely to reduce the scattering on the heterojunction
imperfections and, as a consequence, to decrease the resistivity.

As for the scattering on impurities, it can be shown within the strongly
localized potential approximation
$$U(\vec{r}) = U_0 \,\delta (\vec{\rho} - \vec{\rho}_i)
\,\delta (x - x_i)$$
(here, $\vec{\rho}_i$ is the impurity coordinate in the lateral direction,
and $x$ -- along the $x$-axis), that the incoherent part of the
probability for the electron scattering on such a potential is
proportional to
$$\varphi \sim \int N_S(x) |f_i(x)|^4 dx$$ ($N_S$ is the surface
concentration of impurity),i.e., for a homogeneous distribution of
scattering centers, the scattering probability and, hence, the resistivity
of the nanostructure, are proportional to the integral of the fourth power
of the envelope wave function.

The magnetic field dependence of $\varphi$ value is shown by curve~4
in Fig.~4. It is seen that in magnetic field of the order of 160~kOe
$\varphi$ is more than doubled. It is also seen, that the calculated
curve $\varphi (H)$ is similar to the experimental $R(H)$ dependence,
showing the same tendency to saturation but in higher magnetic field.
(A tendency to saturation of the experimental $R(H)$ curve is seen even
in a field of 50~kOe.) As mentioned above, a decrease in electron
scattering on the heterojunction imperfections with increasing
magnetic field should result in reduced resistivity. If so, the
competition of the two mechanism, when taken into account in the
calculations, may give a better agreement between the calculated and
experimental values of the magnetic field where the saturation begins.

It should be noted in conclusion that the observed transverse
magnetoresistance of this particular asymmetric nanostructure is certain
to have a significant effect on the field dependence of spontaneous PGE
current $J^{PGE}(H)$, but does not explain the nonmonotonic behavior of
$J^{PGE}(H)$ observed in paper~\cite{omel96}.
This nonmonotonic behavior of $J^{PGE}(H)$ is likely to be due to the
PGE nature.

This work was partly supported by the Russian Foundation for Basic
Research, Project No.95-02-04358-a, and partly by Russian Program
``Solid State Nanostructute Physics'', Project No.1-083/4.

\begin{figure}
\caption{TMR measurement circuit and the structure of the
asymmetric system of three GaAs/AlGaAs quantum wells.}
\end{figure}

\begin{figure}
\caption{$J(H)$ dependencies at different bias voltages $-6$~V $< E_v <
6$~V (labeled by the numbers at the curves).}
\end{figure}

\begin{figure}
\caption{Field dependencies of the magnetoresistance at low (1~V) and
high (6~V) absolute values of $E_v$.}
\end{figure}

\begin{figure}
\caption{Calculated magnetic field dependencies of the probabilities
for an electron to be located in the corresponding (according to the
number in Fig.~1) quantum well (curves 1 - 3), and $\varphi$ vs $H$
(curve 4).}
\end{figure}

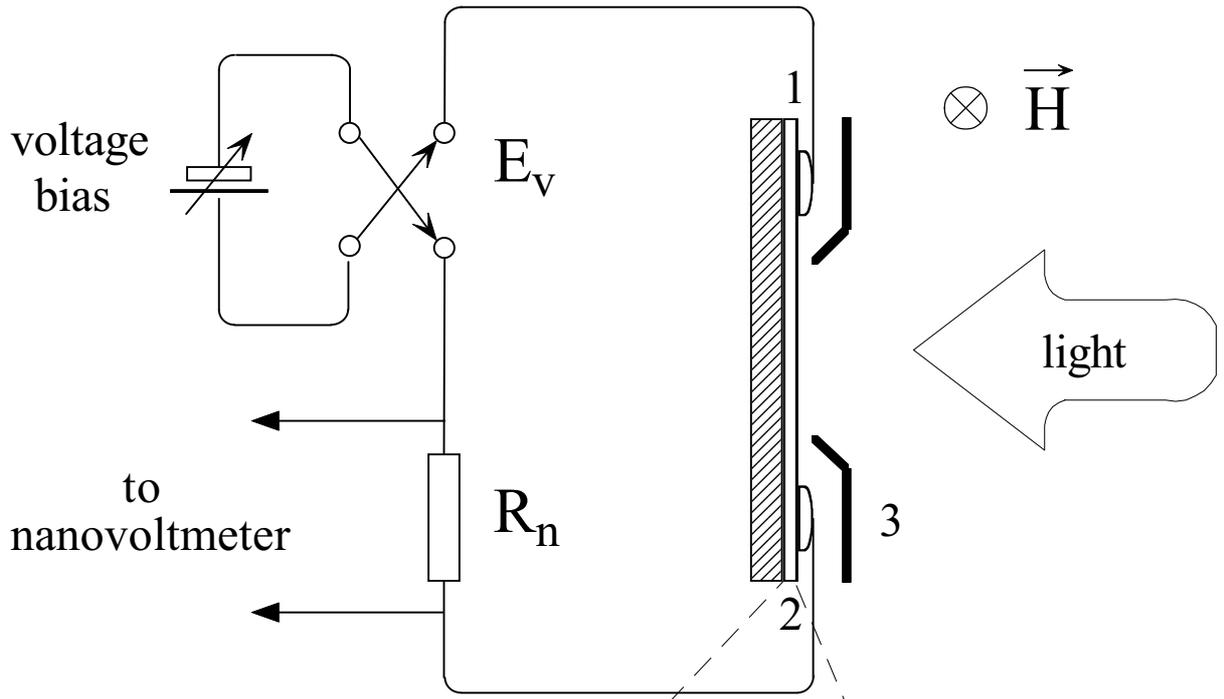
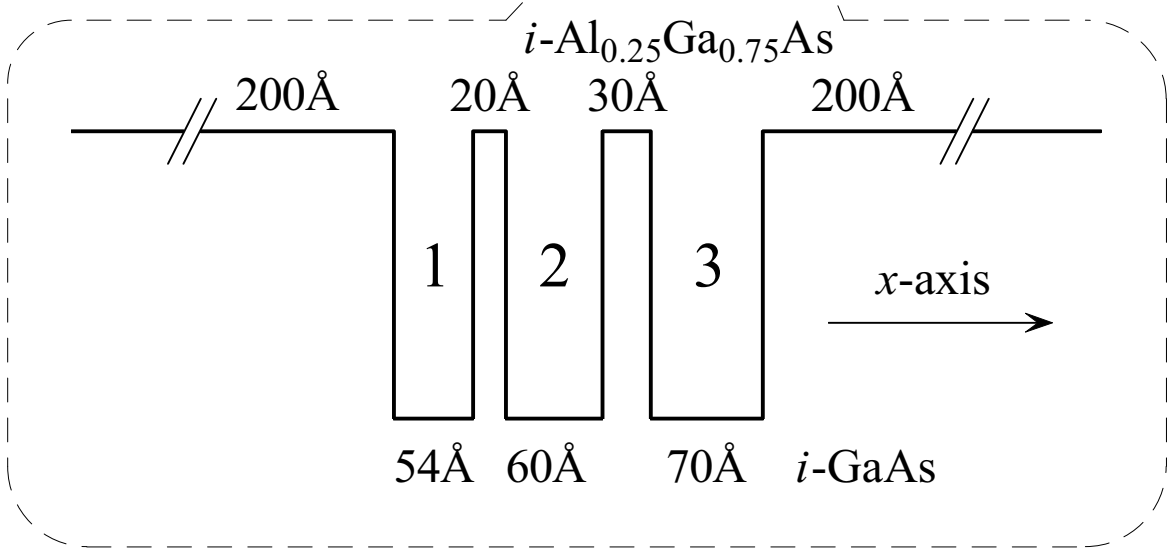
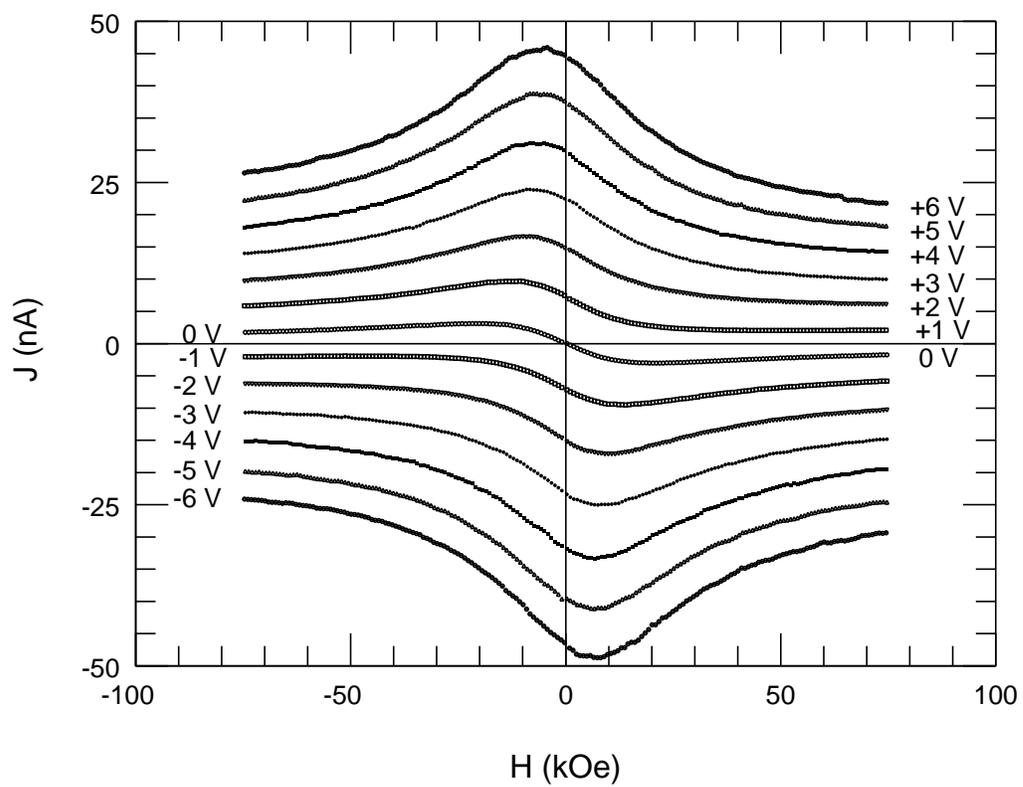
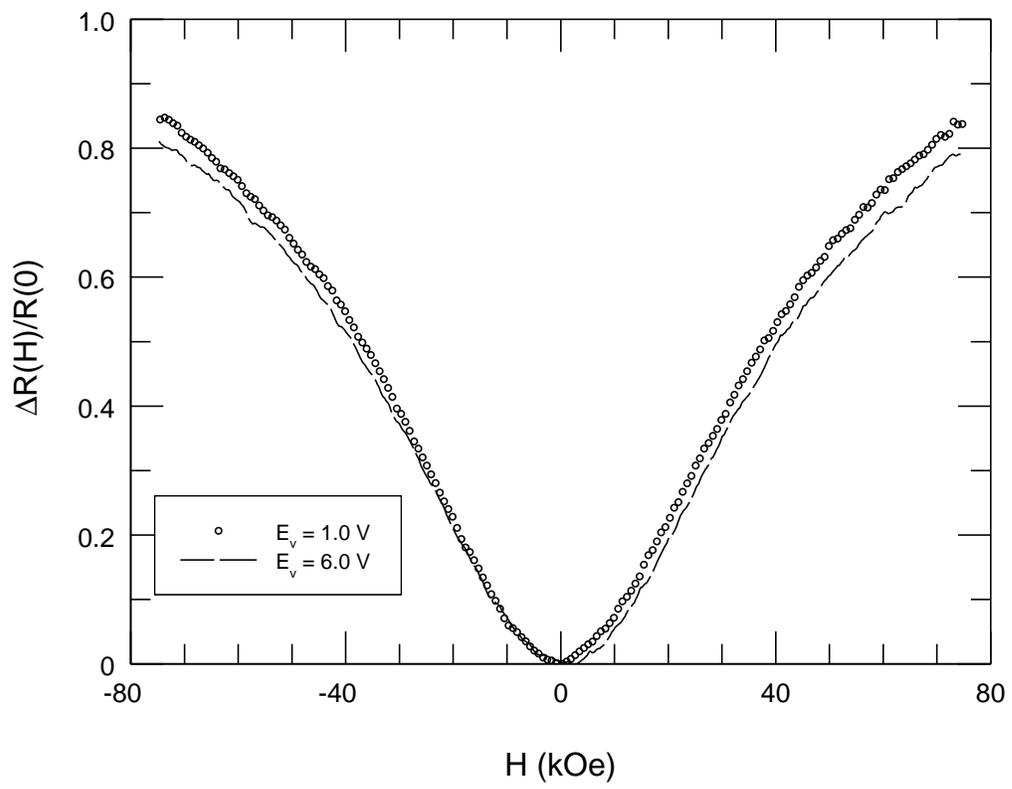
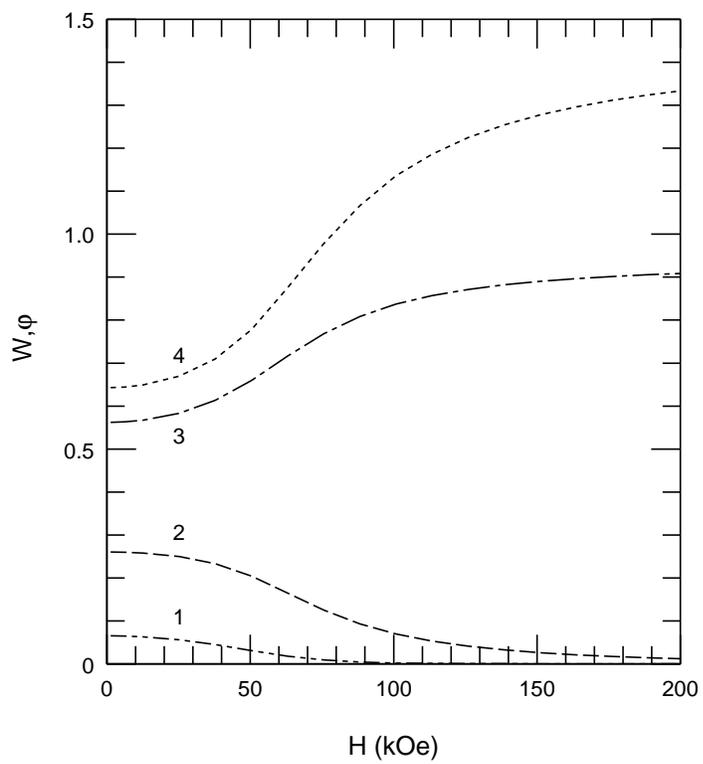
\end{document}